\documentclass[aps,pra,preprintnumbers,showpacs,amsmath,showkeys]{revtex4}

\usepackage{hyperref}
\usepackage{epic,amsmath,graphicx}
\def\mb#1         {\mbox{\boldmath $#1$}}

\begin{document}
\preprint{TKU-07-2 Oct. 2007}

\title{Structure of Mass Gap between Two Spin Multiplets}
\author{Takayuki Matsuki}
\email[E-mail: ]{matsuki@tokyo-kasei.ac.jp}
\affiliation{Tokyo Kasei University,
1-18-1 Kaga, Itabashi, Tokyo 173-8602, JAPAN}
\author{Toshiyuki Morii}
\email[E-mail: ]{morii@kobe-u.ac.jp}
\affiliation{Graduate School of Science and Technology,
Kobe University,\\ Nada, Kobe 657-8501, JAPAN}
\author{Kazutaka Sudoh}
\email[E-mail: ]{kazutaka.sudoh@kek.jp}
\affiliation{Institute of Particle and Nuclear Studies, 
High Energy Accelerator Research Organization, \\ 
1-1 Ooho, Tsukuba, Ibaraki 305-0801, JAPAN}

\date{October 1, 2007}

\begin{abstract}
Studying our semirelativistic potential model and the numerical results, which
succeeds in predicting and reproducing recently discovered higher resonances of $D$,
$D_s$, $B$, and $B_s$, we find a simple expression for the mass gap between two spin
multiplets of heavy-light mesons, $(0^-,1^-)$ and $(0^+,1^+)$. The mass gap
between chiral partners defined
by $\Delta M=M(0^+)-M(0^-)$ and/or $M(1^+)-M(1^-)$ is given by
$\Delta M=M(0^+)-M(0^-)=M(1^+)-M(1^-)\approx \Lambda_{\rm Q}-m_q$ in the limit of heavy
quark symmetry, and including $1/m_Q$ corrections, we have
$\Delta M\approx \Lambda_{\rm Q}-m_q+(1.28\times 10^{5}+4.26\times 10^{2}\cdot m_q)/m_Q$
with $\Lambda_{\rm Q}\approx 300$ MeV, a light quark mass $m_q$, and a heavy quark mass
$m_Q$. This equation holds both for $D$ and $D_s$ heavy mesons. Our model calculations for
the $B$ and $B_s$ also follow this formula.
\end{abstract}
\preprint{}
\maketitle

\section{Introduction}
\label{intro}
The discovery of the narrow $D_{sJ}$ particles by BaBar \cite{BaBar03} and CLEO
\cite{CLEO03} and soon confirmed by Belle \cite{Belle03} immediately reminded
people an effective theory approach proposed by Nowak et al. and others
\cite{Nowak93,Bardeen94,Ebert95,Deandrea98}.  
They constructed an effective Lagrangian for heavy mesons from the
Nambu-Jona-Lasinio type four-fermi interactions and combined it with the chiral
multiplets so that the mass of heavy mesons can be related to the Higgs scalars
of chiral Lagrangian, and they found that two spin multiplets, $j^P=(0^-,1^-)$ and
$(0^+,1^+)$, are degenerate in the limit in which the chiral symmetry is an 
exact symmetry of the vacuum and the heavy quark symmetry is exactly realized.
From this effective theory, they derived the Goldberger-Treiman relation for the mass
gap between chiral partners $0^{+}(1^{+})$ and $0^{-}(1^{-})$ 
instead of the heavy meson mass itself and predicted the mass gap between chiral partners 
of heavy mesons to be around $\Delta M=g_{\pi}f_{\pi}\approx 349$ MeV,
where $g_{\pi}$ is the coupling constant for $0^{+}\rightarrow 0^{-}+\pi$ and $f_{\pi}$
is the pion decay constant.

Finding that the mass gap between chiral partners $0^{+}(1^{+})$ and $0^{-}(1^{-})$ 
in the case of $D_s$ agrees well
with the experiments (around 350 MeV), people thought that underling physics may
be explained by their $SU(3)$ effective Lagrangian \cite{Bardeen03,Harada04}.
However, when $(0^+,1^+)$ for $D$ meson were found by Belle and FOCUS, and later
reanalyzed by CLEO, their explanation needs to be modified even though some people
still study in this direction; in fact, the effective Lagrangian approach \cite{Bardeen03}
predicts about 94 MeV smaller mass gap for $D$ mesons than that for $D_s$ mesons,
while the experimental mass gap for $D$ mesons is about $70\sim 80$ MeV larger than
that for $D_s$ mesons \cite{Belle04}.  Furthermore, what they originally predicted
could not be identified as any of heavy meson multiplets for $D$, $D_s$, $B$, and $B_s$.
In other words, the forumula can be applied equally for any of these heavy meson multiplets. 
Thus, it is required to find the mass gap formula, if it exists, which agrees well with
the experiments and explains the physical ground of its formula.

In this paper, using our semirelativistic potential model, we first give our formula for
the mass gap between chiral partners $0^{+}(1^{+})$ and $0^{-}(1^{-})$ 
for {\it any} heavy meson, $D$, $D_s$, $B$, and $B_s$,
among which the known mass gaps, i.e., the ones for $D$ and $D_s$, agree well with the experiments
although there is some ambiguities for $D$ meson data. Next we show how this mass gap depends
on a light quark mass $m_q$ for $q= u, d$, and $s$, where we neglect the difference between
$u$ and $d$ quarks.
Our formula naturally explain that the mass gap for $D$ is larger than that for $D_s$ and
predict the mass gaps for $B$ and $B_s$.

\section{Semirelativistic Quark Potential Model and Structure of Mass Gap}
\begin{figure}[t]
\includegraphics[scale=1.0,clip]{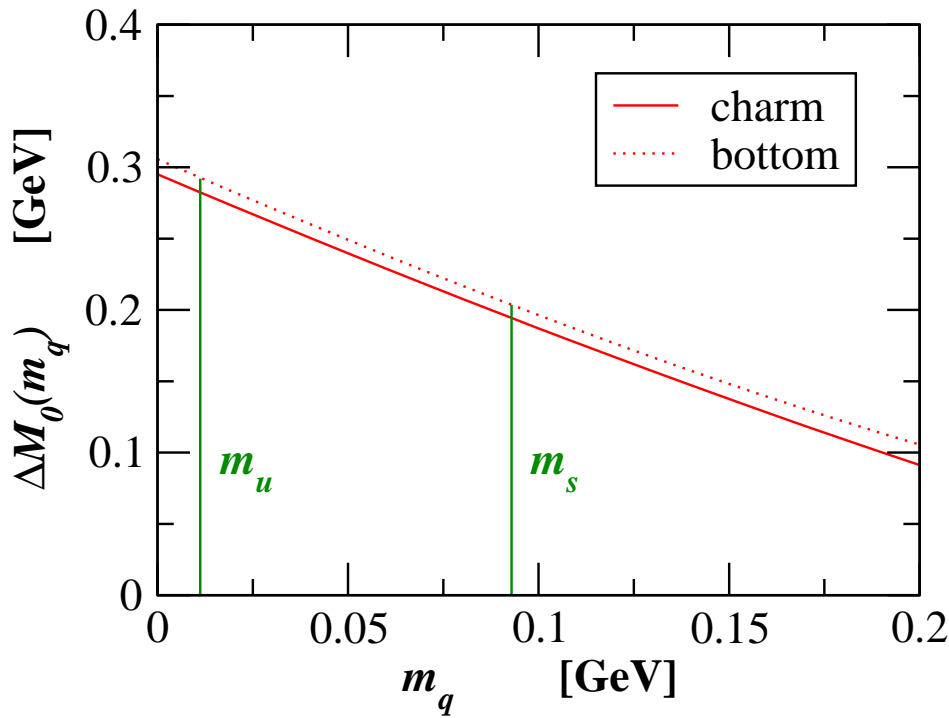}
\caption{Plot of the mass gap between two spin multiplets.
Light quark mass dependence is given. The horizontal axis is light quark mass $m_q$
and the vertical axis is the mass gap $\Delta M_0$.}
\label{fig-DeltaM}
\end{figure}

Mass for the heavy meson $X$ with the spin and parity, $j^P$, is expressed in our formulation
as \cite{Matsuki97}
\begin{eqnarray}
  M_X(j^P) = m_Q + E_0^k(m_q) + O\left(1/m_Q\right), \label{HMass}
\end{eqnarray}
where the quantum number $k$ is related to the total angular momentum $j$ and the
parity $P$ for a heavy meson as \cite{Matsuki04}
\begin{eqnarray}
  j= |k|-1 {\rm ~~or~~} |k|, \quad P=\frac{k}{|k|}(-1)^{|k|+1}, \quad
  E_0^k(m_q)=E_0(j^P, m_q).
\end{eqnarray}
To begin with, we study the heavy meson mass without $1/m_Q$ corrections so that we
can see the essence of the mass gap.
States with the same $|k|$ value are degenerate in a pure chiral limit and without 
confining scalar potential,
which is defined as $m_q\rightarrow 0$ and $S(r) \rightarrow 0$ \cite{Matsuki05}. We consider
the scenario that a chiral symmetry breaking and a confinement take place in two steps. First
the degeneracy is broken due to gluon fields when $S(r)$ is turned on and confines quarks
into heavy mesons but keeping vanishing light quark mass intact.
In fact, in this limit our model gives
the mass gap between two spin multiplets $\Delta M \approx 300$ MeV as follows;
\begin{eqnarray}
  \Delta M=E_0(1^+,0) - E_0(1^-,0)=E_0(0^+,0) - E_0(0^-,0)
  &=& 295.1 {\rm ~MeV~~for~} D, {\rm ~and~} D_s, \nonumber \\
  &=& 309.2 {\rm ~MeV~~for~} B, {\rm ~and~} B_s, 
\end{eqnarray}
This gap is mainly due to gluon fields which confines quarks into heavy mesons.
It is interesting that obtained values are close to
$\Lambda_{\rm QCD}\approx 300~{\rm MeV}$.
Next, turning on a light quark mass
which explicitly breaks a chiral symmetry, we have $SU(3)$ flavor
breaking pattern of the mass levels, i.e., mass of $D$ becomes different from that of
$D_s$ with the same value of $j^P$. Since we assume $m_u=m_d$, there still remains
$SU(2)$ iso-spin symmetry. Note that even after chiral symmetry is broken, there is still
degeneracy between members of a spin multiplet due to the heavy quark symmetry, i.e.,
$SU(2)_f\times SU(2)_{\rm spin}$ symmetry, with $SU(2)_f$ rotational flavor symmetry
and $SU(2)_{\rm spin}$ rotational spin symmetry.
By using the optimal values of parameters in Ref. \cite{Matsuki07}, which is listed
in Table \ref{parameter}, degenerate masses without $1/m_Q$ corrections 
for $D,~D_s$ and $B,~B_s$ mesons are calculated and presented in Table \ref{DegMassgap}.
Furthermore, by changing $m_q$ from 0 to 0.2 GeV, we have calculated the $m_q$
dependence of $\Delta M_0$ and have obtained Fig. \ref{fig-DeltaM}, in which $\Delta M_0$
is linearly decreasing with $m_q$. From Fig. \ref{fig-DeltaM}, we find that the mass gap between 
two spin multiplets for a heavy meson $X$ can be written as
\begin{eqnarray}
  \Delta M_0 &=& M_X(0^+) - M_X(0^-) = M_X(1^+) - M_X(1^-)
  =g_0\Lambda_{\rm Q}-g_1 m_q, \label{DM0} \\
  \Lambda _Q  &=& 300\;{\rm{MeV}},\;
  \left\{ {\begin{array}{*{20}c}
   {g_0  = 0.9836,\;g_1  = 1.080, \quad{\rm for~}D/D_s}  \\
   {g_0  = 1.017,\;g_1  = 1.089, \quad{\rm for~}B/B_s}  \\
  \end{array}} \right.,
\end{eqnarray}
where the values of $g_0$, and $g_1$ are estimated
by fitting the optimal line with Fig. \ref{fig-DeltaM}.
Since both $g_0$ and $g_1$ are very close to 1, we conclude that the mass
gap is essentially given by 
\begin{equation}
\Delta M_0=\Lambda_Q-m_q
\label{MassGap}
\end{equation}
Though the physical ground of this result is out of scope at present, Eq. (\ref{MassGap})
is serious, since it is very different from the one of an effective theory approach 
as mentioned later.  
This result is exact when ${\cal O}\left(1/m_Q\right)$ terms are neglected.
As we will see later,  since $1/m_Q$ corrections are nearly equal
to each other for two spin doublets, the above equation (\ref{MassGap}) between two
spin multiplets holds approximately even with $1/m_Q$ corrections.
%
\begin{table}[t!]
\caption{Optimal values of parameters.}
\label{parameter}
\begin{tabular}{lcccccccc}
\hline
\hline
Parameters
& ~~$\alpha_s^c$ & ~~$\alpha_s^b$ & ~~$a$ (GeV$^{-1}$) & ~~$b$ (GeV) \\
& ~~0.261$\pm$0.001 & ~~0.393$\pm$0.003 & ~~1.939$\pm$0.002 & ~~0.0749$\pm$0.0020 \\
& ~~$m_{u, d}$ (GeV) & ~~$m_s$ (GeV) & ~~$m_c$ (GeV) & ~~$m_b$ (GeV) \\
& ~~0.0112$\pm$0.0019 & ~~0.0929$\pm$0.0021 & ~~1.032$\pm$0.005 & ~~4.639$\pm$0.005 \\
\hline
& \# of data & \# of parameter & total $\chi^2$/d.o.f & \\
& 18 & 8 & 107.55 & \\
\hline
\hline
\end{tabular}
\end{table}

Let us see how the mass gap can be written in our formulation \cite{Matsuki97}. 
Heavy meson mass without $1/m_Q$ corrections can be given by Eq.~(\ref{HMass}) with an eigenvalue $E_0^k$ being given by
the following eigenvalue equation.
%
\begin{equation}
\left( {\begin{array}{*{20}c}
   {m_q  + S + V} & { - \partial _r  + \frac{k}{r}}  \\
   {\partial _r  + \frac{k}{r}} & { - m_q  - S + V}  \\
\end{array}} \right)\left( {\begin{array}{*{20}c}
   {u_k (r)}  \\
   {v_k (r)}  \\
\end{array}} \right) = E_0^k \left( {\begin{array}{*{20}c}
   {u_k (r)}  \\
   {v_k (r)}  \\
\end{array}} \right). \label{eq:app:schrod3}
\end{equation}
Using this equation, the mass gap between $k=+1$ and $k=-1$, which are corresponding to
the spin multiplets $(0^-,1^-)$ and $(0^+,1^+)$, respectively, when they are degenerate,
is re-expressed as
\begin{eqnarray}
  && \Delta M_0 = M^1(0^+) - M^{-1}(0^-) = M^1(1^+) - M^{-1}(1^-)
  \nonumber \\
  &=& \int \frac{d^3x}{4\pi r^2}\left\{\Phi_1^\dagger(r)
\left( {\begin{array}{*{20}c}
   {m_q  + S + V} & { - \partial _r  + \frac{1}{r}}  \\
   {\partial _r  + \frac{1}{r}} & { - m_q  - S + V}  \\
\end{array}} \right)
  \Phi_1(r) -
  {\Phi_{-1}^\dagger(r)
\left( {\begin{array}{*{20}c}
   {m_q  + S + V} & { - \partial _r  - \frac{1}{r}}  \\
   {\partial _r  - \frac{1}{r}} & { - m_q  - S + V}  \\
\end{array}} \right)
  \Phi_{-1}(r)}\right\}
  \nonumber \\
  &=& \int dr \left[\Phi_1^\dagger(r) K_{1} \Phi_1(r)
  - \Phi_{-1}^\dagger(r) K_{-1} \Phi_{-1}(r)\right] + m_q \int dr \left[
  \Phi_1^\dagger(r) \beta \Phi_1(r) -
  \Phi_{-1}^\dagger(r) \beta \Phi_{-1}(r) \right]. \label{Delta}
\end{eqnarray}
From this equation we can see that the mass gap linearly depends on $m_q$.
Here the radial wave function $\Phi_k(r)$ and the massless free kinetic term $K_k$
with the quantum number $k$ are given by
\begin{eqnarray}
  \Phi_k(r) &=&
\left( {\begin{array}{*{20}c}
   {u_k (r)}  \\
   {v_k (r)}  \\
\end{array}} \right),
  \quad
  K_k =
\left( {\begin{array}{*{20}c}
   {S(r) + V(r)} & { - \partial _r  + \frac{k}{r}}  \\
   {\partial _r  + \frac{k}{r}} & { - S(r) + V(r)}  \\
\end{array}} \right).
\end{eqnarray}
Numerically the coefficient of $m_q$ becomes negative, while the first term in
Eq.~(\ref{Delta}) is approximately given by 300 MeV, which is nearly equal to the scale
parameter of QCD, $\Lambda_{\rm QCD}$. That a coefficient of $m_q$ becomes negative in
Eq.~(\ref{Delta}) can be explained or we can intuitively understand this result in our
formulation as follows. The quantum numbers
$k=-1$ and $k=1$ correspond to $\ell=0$ and $\ell=1$ respectively, where $\ell$ is
the angular momentum of a light antiquark relative to a heavy quark as can be seen
from Table I of Ref.\cite{Matsuki07}.
An excited state with $\ell=1$ ($k=+1$) is more relativistic compared with the one
with $\ell=0$ ($k=-1$), which means a lower component $v_1(r)$ becomes larger
than $v_{-1}(r)$ since they are normalized as $(u_k)^2+(v_k)^2=1$. Hence
$(u_1)^2 -(v_1)^2=\Phi_1^\dagger(r) \beta \Phi_1(r)$ becomes smaller
than $\Phi_{-1}^\dagger(r) \beta \Phi_{-1}(r)$. Thus the coefficient of $m_q$ becomes
negative. As a matter of fact, 
linear $m_q$ dependence of $\Delta M$ is not yet definite since radial wave functions
$u_k$ and $v_k$ are also dependent on $m_q$. However, looking at Eq.~(\ref{DM0}) or
Fig.~\ref{fig-DeltaM} which are the numerical calculation of our model, we can say that
implicit dependence on $m_q$ of these wave functions is numerically small. Thus the
above physical and intuitive interpretation of linear $m_q$ dependence of $\Delta M_0$
is correct.
\begin{table}[t!]
\caption{Degenerate masses of model calculations and their mass gap between $0^+(1^+)$
and $0^-(1^-)$ for $n=1$.}
\label{DegMassgap}
\begin{tabular}{lcccc}
\hline
\hline
~~~
& ~~$M_0(D)$ & ~~$M_0(D_s)$ & ~$M_0(B)$ & ~$M_0(B_s)$ \\
\hline
$0^-/1^-$
& ~~1784 & ~~1900 & ~5277 & ~5394 \\
$0^+/1^+$
& ~~2067 & ~~2095 & ~5570 & ~5598 \\
\hline
$0^+(1^+)-0^-(1^-)$
& ~~283 & ~~195 & ~293 & ~204 \\
\hline
\hline
\end{tabular}
\end{table}
%

\section{Interpretation due to Chiral Effective Theory}
The above result suggests that the physical ground of chiral symmetry breakdown or
generation of mass
for heavy mesons occurs differently from what people in \cite{Nowak93,Bardeen94,Ebert95}
originally considered. Let us briefly explain the mechanism that these authors considered as
a generation of the mass gap, which is due to the paper \cite{Bardeen94}.
The Lagrangian for the chiral multiplets, which couples to the heavy quark sector, can be
written as follows.
\begin{eqnarray}
  {\cal L}_{\rm chiral} =\bar\psi\left(i/{\kern -2 mm}\partial -m_q\right)\psi
  -g\bar\psi_L\Sigma\psi_R-g\bar\psi_R\Sigma^\dagger\psi_L
  -\frac{1}{2}\Lambda^2{\rm Tr}\left(\Sigma^\dagger\Sigma\right),
\end{eqnarray}
where $\psi$ is the chiral quark field with three flavors and $\Sigma$ is the $3\times 3$
complex auxiliary field which are given by
\begin{eqnarray}
  \psi^T = \left(u, d, s\right), \quad
  \Sigma = \frac{1}{2}\sigma I_3+i\pi^a\frac{\lambda^a}{2},
\end{eqnarray}
When this Lagrangian is combined
with the effective theory for heavy hadrons, the effective mass of a constituent
quark is given by $\left<\sigma\right>+m_q$. Then the mass gap is given by
\begin{equation}
  \Delta M_0 = g_\pi\left(\left<\sigma\right>+m_q\right) .
\end{equation}
where $g_\pi$ is the Yukawa coupling constant between the heavy meson and a chiral
multiplet and is taken to be $g_\pi=3.73$ in \cite{Bardeen03}, and $\left<\sigma\right>=f_\pi$.
This expression is obtained
in the heavy quark symmetric limit and should be compared with our Eq.~(\ref{MassGap}). Instead
of minus sign for the term $m_q$ that we obtained, the authors of \cite{Bardeen94}
obtained plus sign as shown in the above equation. The same result is obtained even
if we use the nonlinear $\Sigma$ model \cite{Bardeen03}.

\section{$1/m_Q$ Corrections}
\label{correction}
Next let us study the case when $1/m_Q$ corrections to the mass gap are taken into account.
Part of the results is given in \cite{Matsuki06}. In Table \ref{massgap}, we give our
numerical results in the cases of $n=1$ and $n=2$ (radial excitations). Values in brackets
are taken from the experiments. Our values seem to agree with the experimental ones
though the fit is not as good as the case for the absolute values of heavy meson masses.
We assume the form of the mass gap with the $1/m_Q$ corrections as follows.
\begin{equation}
  \Delta M = \Delta M_0 + \frac{c+d\cdot m_q}{m_Q}. \label{DM}
\end{equation}
Using Eq.(\ref{DM0}) for $D$ and $D_s$ mesons, i.e.
$\Delta M_0=g_0\Lambda_{\rm Q}-g_1m_q=295.1-1.080 m_q$, we obtain the values of the
parameters $c$ and $d$ for $D/D_s$ mesons given in Table
\ref{massgap}, which are given by
\begin{equation}
  c = 1.28\times 10^5~{\rm MeV^2}, \quad d = 4.26 \times 10^2~{\rm MeV}. 
  \label{cdParam}
\end{equation}
The term $c/m_Q$ lifts the constant $g_0\Lambda_Q$ about 100 MeV and the term
$d/m_Q$ gives deviation from -1 to the coefficient for $m_q$ in the case of $D/D_s$.

Applying this formula, Eq.~(\ref{DM}), to the case for $B/B_s$ with $m_Q=m_b$,
we obtain the mass gap as follows.
\begin{equation}
  B(0^+)-B(0^-)\approx B(1^+)-B(1^-)\approx 322,\quad
  B_s(0^+)-B_s(0^-)\approx B_s(1^+)-B_s(1^-)\approx 240 \quad {\rm MeV},
\end{equation}
which should be compared with our model calculations, 321 and 241 MeV, in Table
\ref{massgap}. Thus the linear dependence of the mass gap on $m_q$ is also supported
in the case where the $1/m_Q$ corrections are taken into account.  The calculated
$m_q$ dependence  of $\Delta M$ with $1/m_Q$ corrections is presented in Fig. 2,
for $0<m_q<0.2$GeV.
Valuse in Table \ref{massgap} are calculated using those in Tables \ref{DDsmeson} and
\ref{BBsmeson}.
\begin{table}[t!]
\caption{Model calculations of the mass gap. Values in brackets are taken from the
experiments. Units are MeV.}
\label{massgap}
\begin{tabular}{lcccc}
\hline
\hline
Mass gap ($n=1$)
& ~~$\Delta M(D)$ & ~~$\Delta M(D_s)$ & ~$\Delta M(B)$ & ~$\Delta M(B_s)$ \\
\hline
$0^+-0^-$
& ~~414 (441) & ~~358 (348) & ~322 & ~239 \\
$1^+-1^-$
& ~~410 (419) & ~~357 (348) & ~320 & ~242 \\
\hline
\hline
\end{tabular}
\end{table}
\begin{table}[t!]
\begin{tabular}{lcccc}
\hline
\hline
($n=2$)
& ~~$\Delta M(D)$ & ~~$\Delta M(D_s)$ & ~$\Delta M(B)$ & ~$\Delta M(B_s)$ \\
\hline
$0^+-0^-$
& ~~308 & ~~274 & ~~206 & ~~160 \\
$1^+-1^-$
& ~~350 & ~~327 & ~~216 & ~~171 \\
\hline
\hline
\end{tabular}
\end{table}

\begin{figure}[t]
\includegraphics[scale=1.0,clip]{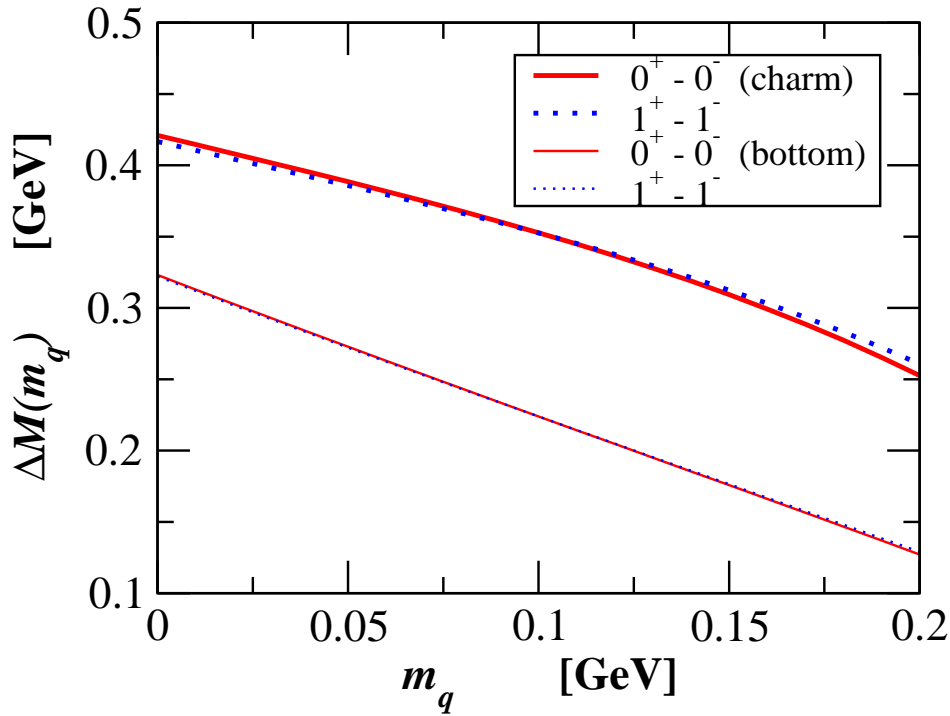}
\caption{Plot of the mass gap between two spin multiplets.
Light quark mass dependence is given. The horizontal axis is light quark mass $m_q$
and the vertical axis is the mass gap $\Delta M$.}
\label{fig2-DeltaM}
\end{figure}

\section{Miscellaneous Phenomena}
{\it Global Flavor $SU(3)$ Recovery -- }
\label{revovery}
Looking at the mass levels of $0^+$ and $1^+$ states for the $D$ and $D_s$ mesons,
one finds that mass differences between $D$ and $D_s$ becomes smaller compared with those
of the $0^-$ and $1^-$ states. This can be seen from Table \ref{DDsmeson} and
was first discussed in Ref.\cite{Dmitrasinovic05} by Dmitra$\check{\rm s}$inovi\'c.
He claimed that considering
$D_{sJ}$ as a four-quark state, one can regard this phenomena as flavor $SU(3)$ recovery.
However, in our interpretation, this is not so as we have seen that this is caused by
the mass gap dependency on a light quark mass, $m_q$, as shown in Fig. \ref{fig-DeltaM}.
That is, when the mass of $D$ meson is elevated largely from the $0^-/1^-$ state to the
$0^+/1^+$ state, the mass of $D_s$ meson is elevated by 
about 100 MeV smaller than that of $0^-/1^-$
as one can see from Fig. \ref{fig-DeltaM}. In our interpretation, the $SU(3)$ is not
recovered since the light quark masses of $m_u=m_d$ and $m_s$ do not change their
magnitudes
when the transition from $0^-/1^-$ to $0^+/1^+$ occurs, and their values remain to be
$m_{u(d)}=11.2$ MeV and $m_s=92.9$ MeV, respectively, as presented in Table \ref{parameter}.

\begin{table*}[t!]
\caption{$D/D_s$ meson mass spectra for both the calculated and experimentally observed ones.
Units are MeV.}
\label{DDsmeson}
\begin{tabular}{@{\hspace{0.5cm}}c@{\hspace{0.5cm}}c
@{\hspace{1cm}}c@{\hspace{0.5cm}}|@{\hspace{0.5cm}}c@{\hspace{1cm}}c@{\hspace{0.5cm}}}
\hline
\hline
$^{2s+1}L_J (J^P)$ &
$M_{\rm calc}(D)$ & $M_{\rm obs}(D)$ &
$M_{\rm calc}(D_s)$ & $M_{\rm obs}(D_s)$ \\
\hline
\multicolumn{1}{@{\hspace{0.6cm}}l}{$^1S_0 (0^-)$} 
& 1869 & 1867
& 1967 & 1969 \\
\multicolumn{1}{@{\hspace{0.6cm}}l}{$^3S_1 (1^-)$} 
& 2011 & 2008
& 2110 & 2112 \\
\multicolumn{1}{@{\hspace{0.6cm}}l}{$^3P_0 (0^+)$} 
& 2283 & 2308
& 2325 & 2317 \\
\multicolumn{1}{@{\hspace{0.6cm}}l}{$"^3P_1" (1^+)$} 
& 2421 & 2427
& 2467 & 2460 \\
\hline
\hline
\end{tabular}
\end{table*}
\begin{table*}[t!]
\caption{$B/B_s$ meson mass spectra for both the calculated and experimentally observed ones.
Units are MeV.}
\label{BBsmeson}
\begin{tabular}{@{\hspace{0.5cm}}c@{\hspace{0.5cm}}c
@{\hspace{1cm}}c@{\hspace{0.5cm}}|@{\hspace{0.5cm}}c@{\hspace{1cm}}c@{\hspace{0.5cm}}}
\hline
\hline
$^{2s+1}L_J (J^P)$ &
$M_{\rm calc}(B)$ & $M_{\rm obs}(B)$ &
$M_{\rm calc}(B_s)$ & $M_{\rm obs}(B_s)$ \\
\hline
\multicolumn{1}{@{\hspace{0.6cm}}l}{$^1S_0 (0^-)$} 
& 5270 & 5279
& 5378 & 5369 \\
\multicolumn{1}{@{\hspace{0.6cm}}l}{$^3S_1 (1^-)$} 
& 5329 & 5325
& 5440 & $-$ \\
\multicolumn{1}{@{\hspace{0.6cm}}l}{$^3P_0 (0^+)$} 
& 5592 & $-$
& 5617 & $-$ \\
\multicolumn{1}{@{\hspace{0.6cm}}l}{$"^3P_1" (1^+)$} 
& 5649 & $-$
& 5682 & $-$ \\
\hline
\hline
\end{tabular}
\end{table*}

{\it Mass Gap of Heavy Baryons --}
\label{baryon}
When we apply our formula to the heavy-light baryons which include two heavy
quarks, $\left(ccs\right)$, $\left(ccu\right)$, $\left(bcs\right)$, $\left(bcu\right)$, $\left(bbs\right)$,
and $\left(bbu\right)$,
mass gaps between two pairs of baryons, like $\left(ccs\right)$ and $\left(ccu\right)$,
will be given by Eq.~(\ref{MassGap}) in the heavy quark symmetric limit and by
Eq.~(\ref{DM}) with $1/m_Q$ corrections where we have to replace $m_Q$ with
$m_{Q_1}+m_{Q_2}$. Here the isospin symmetry is respected since in our model
$m_u=m_d$.
This speculation is legitimized since
$QQ$ pair can be considered to be $3^*$ expression in the color $SU(3)$ space so that the
baryon like $QQq$ can be regarded as a heavy-light meson and our arguments expanded in this
paper can be applied \cite{Savage90,Ito93}.

\def\Journal#1#2#3#4{{#1} {\bf #2}, #3 (#4)}
\def\etal{et al.}
\def\NIM{Nucl. Instrum. Methods}
\def\NIMA{Nucl. Instrum. Methods A}
\def\NPB{Nucl. Phys. B}
\def\PLB{Phys. Lett. B}
\def\PRL{Phys. Rev. Lett.}
\def\PRD{Phys. Rev. D}
\def\PRO{Phys. Rev.}
\def\ZPC{Z. Phys. C}
\def\EPJ{Eur. Phys. J. C}
\def\EPJA{Eur. Phys. J. C}
\def\PR{Phys. Rept.}
\def\IJM{Int. J. Mod. Phys. A}
\def\PTP{Prog. Theor. Phys.}

\end{document}